\documentclass[preprint,12pt,authoryear]{elsarticle}

\usepackage{amssymb}
\usepackage{url}
\usepackage{subfig}
\usepackage{adjustbox}

\journal{SCIENCE CHINA
Information Sciences}

\usepackage{pgfplots}
\usepackage{tikz}
\usetikzlibrary{arrows,shadows,backgrounds,calc,decorations.pathreplacing,decorations.pathmorphing}

\pgfplotscreateplotcyclelist{my black white}{%
solid, every mark/.append style={solid, fill=gray}, mark=*\\%
densely dashdotted,every mark/.append style={solid, fill=gray},mark=diamond*\\%
densely dotted, every mark/.append style={solid, fill=gray}, mark=triangle*\\%
loosely dashed, every mark/.append style={solid, fill=gray},mark=*\\%
dotted, every mark/.append style={solid, fill=gray}, mark=square*\\%
densely dotted, every mark/.append style={solid, fill=gray}, mark=otimes*\\%
dashed, every mark/.append style={solid, fill=gray},mark=diamond*\\%
densely dashed, every mark/.append style={solid, fill=gray},mark=square*\\%
dashdotted, every mark/.append style={solid, fill=gray},mark=otimes*\\%
dasdotdotted, every mark/.append style={solid},mark=star\\%
}

	\definecolor{webgreen}{rgb}{0,.5,0}
	\definecolor{webbrown}{rgb}{.6,0,0}
	\definecolor{webyellow}{rgb}{0.98,0.92,0.73}
	\definecolor{webgray}{rgb}{.753,.753,.753}
	\definecolor{webblue}{rgb}{0,0,.8}
    \definecolor{webgreen}{rgb}{0, 0.7, 0} 
    \definecolor{webred}{rgb}{0.8, 0, 0}   

\pgfplotscreateplotcyclelist{my color list}{%
solid, color=webblue, every mark/.append style={solid, fill=webblue}, mark=*\\%
densely dashdotted, color=webgreen, every mark/.append style={solid, fill=webred},mark=diamond*\\%
densely dotted, color=webbrown, every mark/.append style={solid, fill=webgreen}, mark=triangle*\\%
loosely dashed, color=webred, every mark/.append style={solid, fill=webbrown},mark=*\\%
dotted, color=webblue, every mark/.append style={solid, fill=webyellow}, mark=square*\\%
densely dotted, color=webgreen, every mark/.append style={solid, fill=gray}, mark=otimes*\\%
dashed, color=webbrown, every mark/.append style={solid, fill=gray},mark=diamond*\\%
densely dashed, every mark/.append style={solid, fill=gray},mark=square*\\%
dashdotted, every mark/.append style={solid, fill=gray},mark=otimes*\\%
dasdotdotted, every mark/.append style={solid},mark=star\\%
}

\begin{document}

\begin{frontmatter}



\title{Comments on the parallelization efficiency\\of the Sunway TaihuLight supercomputer}

\author[label1]{J\'anos V\'egh}
\ead{J.Vegh@uni-miskolc.hu}

\address[label1]{University of Miskolc, Hungary}

\begin{abstract}

In the world of supercomputers, the large number of processors 
requires to minimize the inefficiencies of parallelization, which
appear as a sequential part of the program from the point of view
of Amdahl's law. The recently suggested new figure of merit is applied
to the recently presented supercomputer, and the timeline of "Top 500"
supercomputers is scrutinized using the metric. It is demonstrated,
that in addition to the computing performance and power consumption,
the new supercomputer is also excellent in the efficiency of parallelization.
Based on the suggested merit, a  "Moore-law" like observation is derived
for the timeline of parallelization efficacy  of supercomputers.

\end{abstract}

\begin{keyword}
	supercomputer, parallelization, performance, scaling, figure of merit
\end{keyword}

\end{frontmatter}


\section{Introduction}\label{sec:introduction}

\noindent 
Supercomputers are ranked~(\cite{Top500:2016}) according to their parameter "Rmax (TFlop/s)",
which parameter depends of two factors: how many processors are comprised and how effectively
they are put together. Increasing the number of processors only is useless, as pointed out early by Amdahl:~(\cite{AmdahlSingleProcessor67})
\emph{the effort expended
	on achieving high parallel processing rates is wasted unless it is accompanied by achievements in
	sequential processing rates of very nearly the same magnitude}".
	
Most of the users of supercomputers are not using all available processors, they are rather interested in 
the efficiency of parallelization of their program.
To find a proper merit was always subject of serious debates~(see~\cite{Sun:BetterPerformanceMetric1991}).
It looks like the recently introduced figure of merit~(see~\cite{VeghAlphaEff:2016}), the \textit{effective parallelization},
is a good merit not only to characterize the effectivity of parallelizing software execution, but also 
to characterize  the engineering ingenuity of parallelizing the hardware operation, and so allows to characterize the timeline
of supercomputer development itself.

 \section{\uppercase{The merit $\alpha_{eff}$}}
 
 According to Amdahl~(\cite{AmdahlSingleProcessor67}), the speedup can be expressed as
 
 \begin{equation}
 S^{-1}=(1-\alpha) +\alpha/k
 \end{equation}
 
 \noindent where $k$ is the number of parallelized processors, 
 $\alpha$ is the ratio of the parallelizable part  to the total sequential part,
 $S$ is the measurable speedup. The same relation can be expressed~(see~\cite{VeghAlphaEff:2016}) also in the form
 
 \begin{equation}
 \alpha_{eff} = \frac{k}{k-1}\frac{S-1}{S}
 \end{equation}
 
 The first form is an architectural view, the second one is empirical: no matter,
 \textit{what} causes the (apparently) sequential part, ($1-\alpha$)  part decreases the parallelism, and so can be used to quantitize the goodness of the implementation of parallelisation.
 
 In general, the \textit{efficiency} (in the case of supercomputers: $\frac{R_{max}}{R_{peak}}$) is used, which cannot be used as a single parameter to describe 
 the efficacy of the implementation.
 When using several processors, one of them makes the sequential calculation, the others are waiting
 (use the same amount of time). So, when calculating the speedup, one calculates
 
 \begin{equation}
 S=\frac{(1-\alpha)+\alpha}{(1-\alpha)+\alpha/k} =\frac{k}{k(1-\alpha)+\alpha}
 \end{equation}
 hence the  efficiency
 \begin{equation}
 R = \frac{S}{k}=\frac{1}{k(1-\alpha)+\alpha}\label{eq:soverk}
 \end{equation}
 
 This explains the behavior of diagram $\frac{S}{k}$ in function of $k$:
 the more processors, the lower efficiency, and the larger $(1-\alpha)$,
 the lower is the reachable speedup.
 
 At this point one can notice that $\frac{1}{R}$ is a linear function of the number of the processors, and its slope equals to $(1-\alpha)$, i.e. from the speedup data
 one can estimate value of $\alpha$ even for the individual regions,
 i.e. without knowing the execution time on 1 processor
 (from technical reasons, it is the usual case in the case of supercomputers).

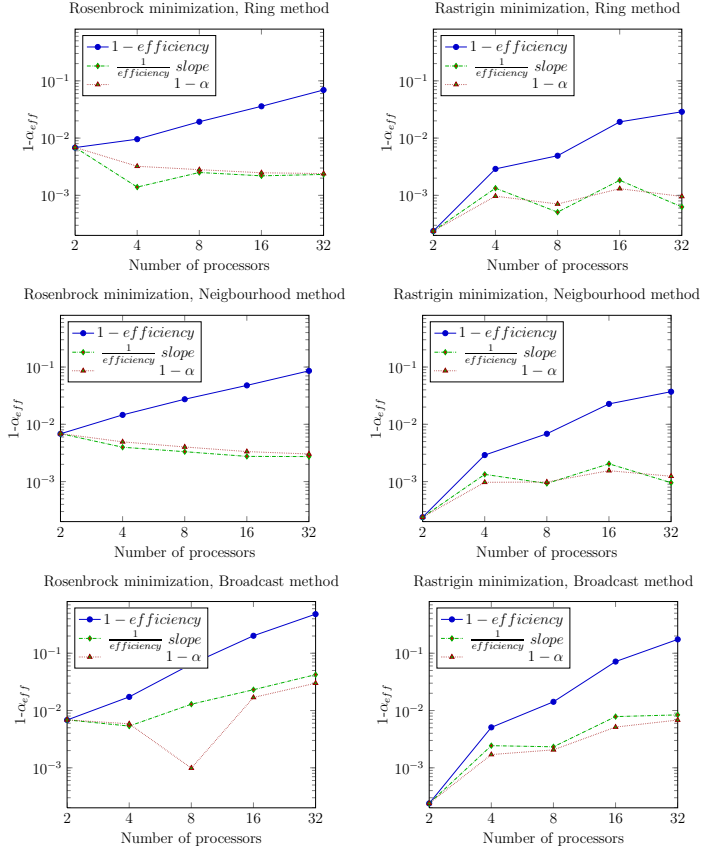
\begin{figure*}
\maxsizebox{.7\textwidth}{!}
{\centering
\begin{tabular}{cc}
\begin{tikzpicture}
\begin{axis}
[
	cycle list name={my color list},
	title={Rosenbrock minimization, Ring method},
		legend style={
			cells={anchor=east},
			legend pos={north west},
		},
		xmin=2, xmax=32,
		ymin=2.e-4, ymax=8.e-1, 
		xlabel=Number of processors,
		ylabel={1-$\alpha_{eff}$},
		ymode=log,
		log basis x=2,
		xmode=log,
		xticklabel=\pgfmathparse{2^\tick}\pgfmathprintnumber{\pgfmathresult}
		]
\addplot table [x=a, y=c, col sep=comma] {RosenbrockRing.csv};
		\addlegendentry{$1-efficiency$}
\addplot table [ y=d, col sep=comma] {RosenbrockRing.csv};
		\addlegendentry{$\frac{1}{efficiency}\ slope$}
\addplot table [ y=e, col sep=comma] {RosenbrockRing.csv};
		\addlegendentry{$1-\alpha$}
\end{axis}
\end{tikzpicture}
&
\begin{tikzpicture}
\begin{axis}
[
	title={Rastrigin minimization, Ring method},
	cycle list name={my color list},
		legend style={
			cells={anchor=east},
			legend pos={north west},
		},
		xmin=2, xmax=32,
		ymin=2.e-4, ymax=8.e-1, 
		xlabel=Number of processors,
		ylabel={1-$\alpha_{eff}$},
		ymode=log,
		log basis x=2,
		xmode=log,
		xticklabel=\pgfmathparse{2^\tick}\pgfmathprintnumber{\pgfmathresult}
		]
\addplot table [x=a, y=c, col sep=comma] {RastriginRing.csv};
		\addlegendentry{$1-efficiency$}
\addplot table [ y=d, col sep=comma] {RastriginRing.csv};
		\addlegendentry{$\frac{1}{efficiency}\ slope$}
\addplot table [ y=e, col sep=comma] {RastriginRing.csv};
		\addlegendentry{$1-\alpha$}
\end{axis}
\end{tikzpicture}
\\
\begin{tikzpicture}
\begin{axis}
[
	title={Rosenbrock minimization, Neigbourhood method},
	cycle list name={my color list},
		legend style={
			cells={anchor=east},
			legend pos={north west},
		},
		xmin=2, xmax=32,
		ymin=2.e-4, ymax=8.e-1, 
		xlabel=Number of processors,
		ylabel={1-$\alpha_{eff}$},
		ymode=log,
		log basis x=2,
		xmode=log,
		xticklabel=\pgfmathparse{2^\tick}\pgfmathprintnumber{\pgfmathresult}
		]
\addplot table [x=a, y=c, col sep=comma] {RosenbrockNeighbour.csv};
		\addlegendentry{$1-efficiency$}
\addplot table [ y=d, col sep=comma] {RosenbrockNeighbour.csv};
		\addlegendentry{$\frac{1}{efficiency}\ slope$}
\addplot table [ y=e, col sep=comma] {RosenbrockNeighbour.csv};
		\addlegendentry{$1-\alpha$}
\end{axis}
\end{tikzpicture}
&
\begin{tikzpicture}
\begin{axis}
[
	title={Rastrigin minimization, Neigbourhood method},
	cycle list name={my color list},
		legend style={
			cells={anchor=east},
			legend pos={north west},
		},
		xmin=2, xmax=32,
		ymin=2.e-4, ymax=8.e-1, 
		xlabel=Number of processors,
		ylabel={1-$\alpha_{eff}$},
		ymode=log,
		log basis x=2,
		xmode=log,
		xticklabel=\pgfmathparse{2^\tick}\pgfmathprintnumber{\pgfmathresult}
		]
\addplot table [x=a, y=c, col sep=comma] {RastriginNeighbour.csv};
		\addlegendentry{$1-efficiency$}
\addplot table [ y=d, col sep=comma] {RastriginNeighbour.csv};
		\addlegendentry{$\frac{1}{efficiency}\ slope$}
\addplot table [ y=e, col sep=comma] {RastriginNeighbour.csv};
		\addlegendentry{$1-\alpha$}
\end{axis}
\end{tikzpicture}
\\
\begin{tikzpicture}
\begin{axis}
[
	title={Rosenbrock minimization, Broadcast method},
	cycle list name={my color list},
		legend style={
			cells={anchor=east},
			legend pos={north west},
		},
		xmin=2, xmax=32,
		ymin=2.e-4, ymax=8.e-1, 
		xlabel=Number of processors,
		ylabel={1-$\alpha_{eff}$},
		ymode=log,
		log basis x=2,
		xmode=log,
		xticklabel=\pgfmathparse{2^\tick}\pgfmathprintnumber{\pgfmathresult}
		]
\addplot table [x=a, y=c, col sep=comma] {RosenbrockBroadcast.csv};
		\addlegendentry{$1-efficiency$}
\addplot table [ y=d, col sep=comma] {RosenbrockBroadcast.csv};
		\addlegendentry{$\frac{1}{efficiency}\ slope$}
\addplot table [ y=e, col sep=comma] {RosenbrockBroadcast.csv};
		\addlegendentry{$1-\alpha$}
\end{axis}
\end{tikzpicture}
&
\begin{tikzpicture}
\begin{axis}
[
	title={Rastrigin minimization, Broadcast method},
	cycle list name={my color list},
		legend style={
			cells={anchor=east},
			legend pos={north west},
		},
		xmin=2, xmax=32,
		ymin=2.e-4, ymax=8.e-1, 
		xlabel=Number of processors,
		ylabel={1-$\alpha_{eff}$},
		ymode=log,
		log basis x=2,
		xmode=log,
		xticklabel=\pgfmathparse{2^\tick}\pgfmathprintnumber{\pgfmathresult}
		]
\addplot table [x=a, y=c, col sep=comma] {RastriginBroadcast.csv};
		\addlegendentry{$1-efficiency$}
\addplot table [ y=d, col sep=comma] {RastriginBroadcast.csv};
		\addlegendentry{$\frac{1}{efficiency}\ slope$}
\addplot table [ y=e, col sep=comma] {RastriginBroadcast.csv};
		\addlegendentry{$1-\alpha$}
\end{axis}
\end{tikzpicture}
\\
\end{tabular}
} 
\caption{Comparing efficiency, efficiency slope and $\alpha_{eff}$
for different communication strategies when running two minimization task on SoC by~\cite{ReconfigurableAdaptive2016}}
\label{fig:SoCCommunicationDiagram}
\end{figure*}

Notice also that through using Equ.~(\ref{eq:soverk}), $\frac{S}{k}$ can be equally good for describing the efficiency of parellelization efficiency of a setup,
if the number of processors is also known. From Equ. (\ref{eq:soverk})
\begin{equation}
\alpha_{R} = \frac{R k -1}{R (k-1)}\label{eq:alphafromr}
\end{equation}


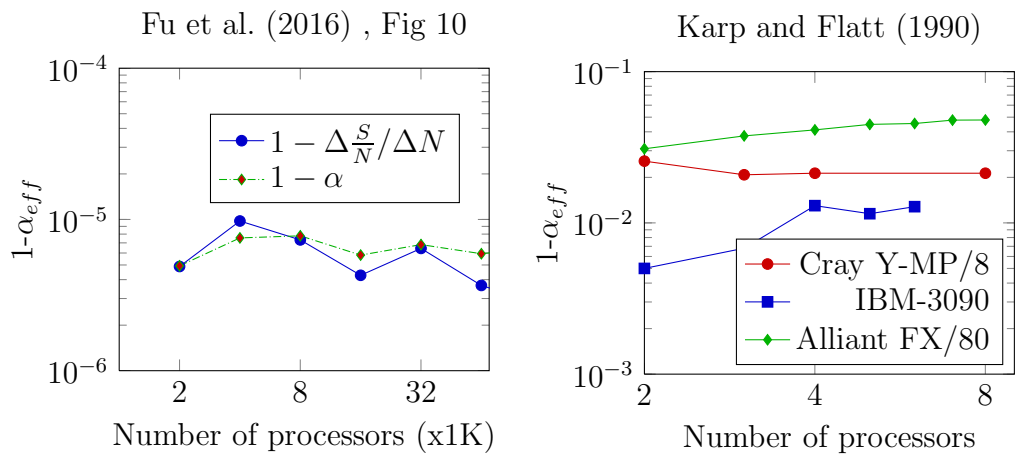
\begin{figure}[]
	\pgfplotsset{width=6.5cm}
	\centering
	\subfloat
	{
\begin{tikzpicture}
\begin{axis}
[
	cycle list name={my color list},
	title={ \protect{\cite{FuSunwaySystem2016}} , Fig 10},
		legend style={
			cells={anchor=west},
			at={(axis cs:50,5e-5)}
		},
		xmin=1, xmax=70,
		ymin=1.e-6, ymax=1e-4, 
		xlabel=Number of processors (x1K),
		ylabel={1-$\alpha_{eff}$},
		ymode=log,
		log basis x=2,
		xmode=log,
		xticklabel=\pgfmathparse{2^\tick} \pgfmathprintnumber{\pgfmathresult}
		]
\addplot table [ y=d, col sep=comma] {SunwayFig10.csv};
		\addlegendentry{$1-\Delta \frac{S}{N}/\Delta N$}
\addplot table [ y=e, col sep=comma] {SunwayFig10.csv};
		\addlegendentry{$1-\alpha$}
\end{axis}
\end{tikzpicture}
} \subfloat
{
		\begin{tikzpicture}
		\begin{axis}[
   	title={ \protect{\cite{Karp:parallelperformance1990}}},
			/pgf/number format/.cd,
		use comma,
		1000 sep={},
		legend style={
			cells={anchor=east},
			legend pos=south east,
		},
		xmin=2, xmax=9,
		ymin=.001, ymax=0.1, 
		xlabel=Number of processors,
		ylabel={1-$\alpha_{eff}$},
		ymode=log,
		xmode=log,
		log basis x=2,
		xticklabel=\pgfmathparse{2^\tick} \pgfmathprintnumber{\pgfmathresult},
	cycle list name={my color list},
		]
		\addplot[  color=webred,mark=*
		] plot coordinates {
			(2,0.0256)  
			(3,0.0208) 
			(4,0.0213)
			(8,0.0213) 
		};
		\addlegendentry{Cray Y-MP/8}
		\addplot[  color=webblue,mark=square*
		] plot coordinates {
			(2, 0.0050) 
			(3, 0.0068)
			(4, 0.0130)
			(5, 0.0115)
			(6, 0.0128)
		};
		\addlegendentry{IBM-3090}
		\addplot[  color=webgreen,mark=diamond*
		] plot coordinates {
			(2 ,0.0309)  
			(3, 0.0376) 
			(4, 0.0412)
			(5, 0.0448) 
			(6, 0.0454) 
			(7, 0.0478)
			(8, 0.0479) 
		};
		\addlegendentry{Alliant FX/80}
		
		\end{axis}
		\end{tikzpicture}
	}

	\caption{($1-\alpha_{eff}$) values for running benchmark Linpack on Sunway TaihuLight supercomputer and supercomputers 25 years ago,  with different number of parallel
	processors. \protect{\cite{Karp:parallelperformance1990}}
		\label{fig:KarpDiagram1}}
\end{figure}

\noindent This quantity of course assumes that $\alpha$ is independent
from the number of the processors. Its numerical value equals to the value
calculated using differences over the full range of processors, and so 
is not displayed in Fig.~\ref{fig:SoCCommunicationDiagram}. 
The supercomputer technology, according to the need mentioned above,
is focussing on decreasing the (apparently) sequential part
($1-\alpha$), so this quantity is shown on the diagrams rather than $\alpha$ itself.

\section{Characterizing effect of communication method in SOC}

As mentioned, in the Amdahl's model there are only two categories: everything which does not make useful computational work, but needs time,
contributes to the sequential part. Such contribution is 
the internal communication between cores inside a chip. 
In their work ~\cite{ReconfigurableAdaptive2016} compare the effect of using different
internal communication methods. From their speedup results, 
the diagrams shown in Fig.~\ref{fig:SoCCommunicationDiagram} were derived. The diagrams show ($1-\alpha$), and for comparison,
the slope of $\frac{1}{R}$ is also displayed. It looks like
within the limits of the experimental precision, both methods provide the same numerical value.
Also displayed for comparison the diagram $(1-\frac{S}{k})$,
which is traditionally used to describe the performance of multi-processor systems.
As shown, for very low number of processors, the diagram practically
provides the same numerical value,  so it is as good for describing multiprocessor efficiency, as ($1-\alpha$).
However, $(1-\frac{S}{k})$ steadily raises with increasing the processor numbers; in the region typical for supercomputers, is not usable any more.

\section{Characterizing supercomputer architecture}

In supercomputers, the "sequential part" is technically of different origin, but has the same effect on ($1-\alpha_{eff}$).
 The recent chinese supercomputer~(\cite{FuSunwaySystem2016})
 provided also performance data, from which diagrams on Fig~\ref{fig:KarpDiagram1} were derived.
Compare these values (and consider the different scales!)
to the former supercomputer data~(\cite{Karp:parallelperformance1990})  shown in Fig~\ref{fig:KarpDiagram1}; the change is imposant.
The new chinese supercomputer is not only good in energy consumption, and the raw computing power,
but also the coordination of the parallel work is excellently organized (the scale is the same as in Fig.~\ref{fig:SoCCommunicationDiagram}, where inside-chip organization takes place, although there the benchmark is different).

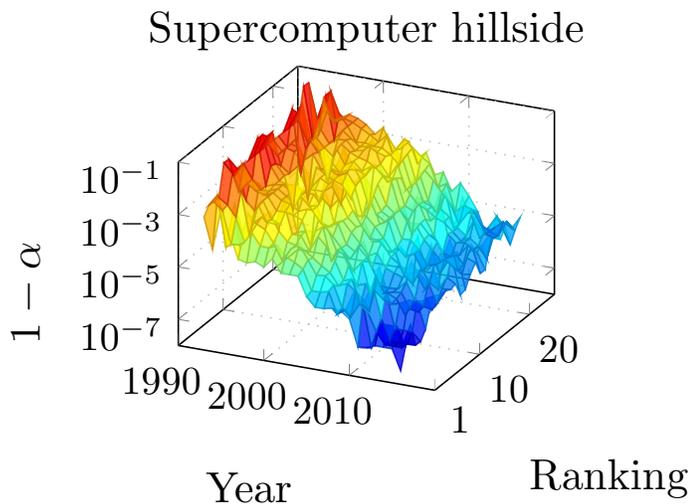
\begin{figure*}
\begin{tikzpicture}[scale=1.5]
\pgfplotstableset{%
    col sep=semicolon,
    z index=0,
    y index=1,
    x index=2,
    header=false
}%
\pgfplotsset{
    every axis/.append style={
        scale only axis,
        width=\textwidth,
       height=\textwidth,
        xtick={1990,2000,2010},
        ytick={1,10,20},
        ztick={1e-7,1e-5,1e-3,1e-1}
    },
    /tikz/every picture/.append style={
        trim axis left,
        trim axis right,
    }
    }
   \begin{axis}
      [ 
    domain=1980:2020,
    domain y=1:40,
      footnotesize,
      title={Supercomputer hillside},
			/pgf/number format/.cd,
		use comma,
		1000 sep={},
		xlabel=Year,
		ylabel=Ranking,
		zlabel=$1-\alpha$,
		xmin=1990, xmax=2020,
		ymin=1, ymax=25, 
		zmin=1e-8, zmax=1e-1, 
		xlabel=Year,
		zmode = log,
		mesh/cols=25,
		grid=major,
		grid style={dotted},
		colormap/jet,
        zmode=log,
      ]

        \addplot3[%
            surf,
            opacity=0.8
        ] table {SupercomputerHillsideNew.csv};

\end{axis}
\end{tikzpicture}
\caption{Supercomputer parallelization efficiency, in function of time and ranking}
\label{SupercomputerHillside}
\end{figure*}

\section{Characterizing the supercomputer timeline}

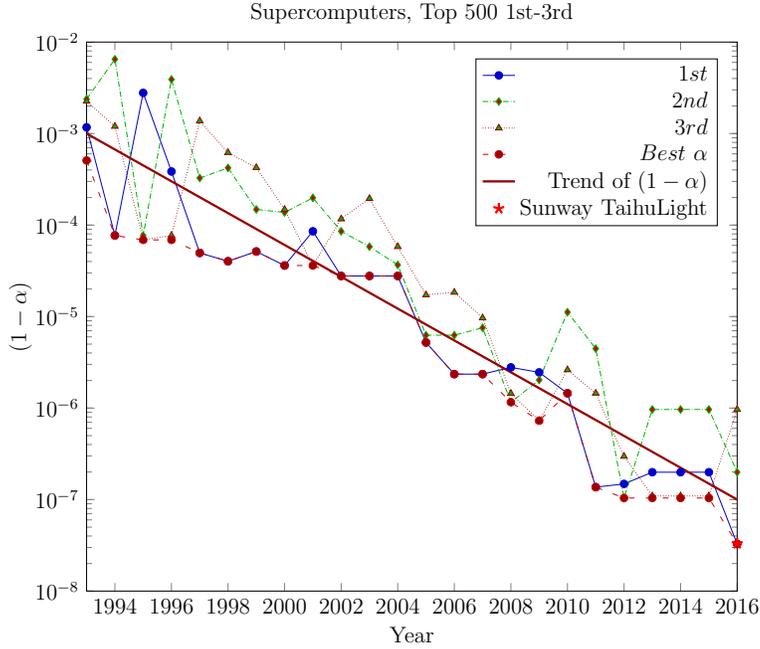
\begin{figure*}

\pgfplotscreateplotcyclelist{my black white}{%
solid, 
every mark/.append style={solid, fill=gray}, mark=*\\%
densely dashdotted,every mark/.append style={solid, fill=gray},mark=diamond*\\%
densely dotted, every mark/.append style={solid, fill=gray}, mark=triangle*\\%
loosely dashed, every mark/.append style={solid, fill=gray},mark=*\\%
dotted, every mark/.append style={solid, fill=gray}, mark=square*\\%
densely dotted, every mark/.append style={solid, fill=gray}, mark=otimes*\\%
dashed, every mark/.append style={solid, fill=gray},mark=diamond*\\%
densely dashed, every mark/.append style={solid, fill=gray},mark=square*\\%
dashdotted, every mark/.append style={solid, fill=gray},mark=otimes*\\%
dasdotdotted, every mark/.append style={solid},mark=star\\%
}

\maxsizebox{.8\textwidth}{\textheight}
{
\begin{tabular}{cc}
\begin{tikzpicture}
\begin{axis}
[
	title={Supercomputers, Top 500 1st-3rd},
	width=\textwidth,
	cycle list name={my color list},
		legend style={
			cells={anchor=east},
			legend pos={north east},
		},
		xmin=1993, xmax=2016,
		ymin=1e-8, ymax=1e-2, 
		xlabel=Year,
		/pgf/number format/1000 sep={},
		ylabel=$(1-\alpha)$,
		ymode=log,
		log basis x=2,
		]
\addplot table [x=a, y=b, col sep=comma] {Top500-0.csv};
		\addlegendentry{$1st $}
\addplot table [x=a, y=c, col sep=comma] {Top500-0.csv};
		\addlegendentry{$2nd $}
\addplot table [x=a, y=d, col sep=comma] {Top500-0.csv};
		\addlegendentry{$3rd $}
\addplot table [x=a, y=e, col sep=comma] {Top500-0.csv};
		\addlegendentry{$Best\ \alpha$}
		\addplot[ very thick, color=webbrown] plot coordinates {
			(1993, 1e-3)  
			(2016,1e-7) 
		};
		\addlegendentry{Trend of $(1-\alpha)$}
		\addplot[only marks, color=red, mark=star,  mark size=3, very thick] plot coordinates {
			(2016,33e-9) 
		};
		\addlegendentry{Sunway TaihuLight}
\end{axis}
\end{tikzpicture}
&
\\
\end{tabular}
} 
\caption{Timeline of supercomputer parallelism}
\label{SupercomputerTimeline}
\end{figure*}

When comparing the performance scales one sees an imposant 
change in the performance.   
There are (not fully detailed) data available on site~\cite{Top500:2016},  covering the "supercomputer age", so
using the data  $R_{max}$ and $R_{peak}$,  and using Equ.~(\ref{eq:alphafromr}), 
($1-\alpha$) can be calculated in function of time and ranking, see Fig~\ref{SupercomputerHillside}. It looks like ($1-\alpha$)
changes in an exponential-like way, both with the time and the ranking in a given year. To establish a more quantitative description, it is worth to derive
a timeline for the past 24 years. In Fig.~\ref{SupercomputerTimeline}, the $(1-\alpha)$ values
are displayed, for the top 3 supercomputers, in function of the time.
The figure also contains the diagram of the best $(1-\alpha)$ in the year,
which confirms that high computing performance strongly correlates with 
the efficiency of parallelization.
It looks like this development path
(independently of technology, manufacturer, number and type of processors) shows a semi-logarithmic behavior,
and only part of the tendency is caused by the Moore-observation.
It is able to forecast the expected behavior of performance in the coming years,
and its validity can provoke debates like the Moore observation does.

\section{Conclusions}
\label{sec:conclusion}

The recently introduced figure of merit "effective parallelization" 
can excellently used to characterize  the quality of hardware 
implementation, too. In addition to qualifying 
manual or compiler optimized parallelization, it can qualify the effect of method of inter-core communication in SoC, can characterize "goodness" of supercomputer implementation. Since a single 
figure of merit describing their performance can be attached to the supercomputers, the timeline of the development of
supercomputing technology can be described. Interestingly enough, 
the timeline of the introduced parameters follow a tendency, similar to the Moore "law".

\section*{References}


  \bibliographystyle{elsarticle-harv} 
  \bibliography{ParallelTOPC}


%
%
%
\end{document}